\documentclass[prd,twocolumn,superscriptaddress]{revtex4}

\newcommand{\bw}{\begin{widetext}}
\newcommand{\ew}{\end{widetext}}

\usepackage{graphicx}
\usepackage{amssymb,amsmath,amsbsy}
\usepackage{hyperref}

\newcommand{\be}{\begin{equation}}
\newcommand{\ee}{\end{equation}}
\newcommand{\bea}{\begin{eqnarray}}
\newcommand{\eea}{\end{eqnarray}}

\newcommand{\bracket}[2]{\bra{#1}\,#2\rangle} 
\newcommand{\bra}[1]{\langle\,#1\,|}          
\newcommand{\ket}[1]{|\,#1\,\rangle}          

\newcommand{\ud}{\mathrm{d}}

\newcommand{\A}{\boldsymbol{A}}

\newcommand{\e}{\mathrm{e}}

\newcommand{\x}{\boldsymbol{x}}

\newcommand{\z}{\boldsymbol{z}}

\newcommand{\hbo}{\hbox to 1 true cm {\hfill } }
\newcommand{\tr}{\hbox{tr}}

\begin{document} 
\title{Is the ground state of Yang--Mills theory Coulombic?}
\author{T.~Heinzl}\email[]{theinzl@plymouth.ac.uk}
\author{A.~Ilderton}\email[]{abilderton@plymouth.ac.uk}
\author{K.~Langfeld}\email[]{klangfeld@plymouth.ac.uk}
\author{M.~Lavelle}\email[]{mlavelle@plymouth.ac.uk}
\affiliation{School of Mathematics \& Statistics, University of Plymouth, Plymouth, PL4 8AA, UK}
\author{W.~Lutz}\email[]{wlutz@plymouth.ac.uk}
\affiliation{School of Mathematics \& Statistics, University of Plymouth, Plymouth, PL4 8AA, UK}
\affiliation{Institut f\"ur Theoretische Physik, Universit\"at T\"ubingen, Auf der Morgenstelle 14, D-72076 T\"ubingen, Germany} 
\author{D.~McMullan}\email[]{dmcmullan@plymouth.ac.uk}
\affiliation{School of Mathematics \& Statistics, University of Plymouth, Plymouth, PL4 8AA, UK}

\begin{abstract}
We study trial states modelling the heavy quark--antiquark ground state in SU(2) Yang--Mills theory. A state describing the flux tube between quarks as a thin string of glue is found to be a poor description of the continuum ground state; the infinitesimal thickness of the string leads to UV artifacts which suppress the overlap with the ground state. Contrastingly, a state which surrounds the quarks with non--abelian Coulomb fields is found to have a good overlap with the ground state for all charge separations. In fact, the overlap increases as the lattice regulator is removed. This opens up the possibility that the Coulomb state is the true ground state in the continuum limit.
\end{abstract}

\maketitle
\section{Introduction}\label{intro}
Explaining colour confinement remains the outstanding problem in the theory of strong interactions. The confinement mechanism is typically pictured, in the mesonic sector for example, as resulting from a change in the distribution of the glue around two quarks as they are moved apart. It is expected that the glue forms a flux tube, the exact geometry of which is unknown, connecting the charges and resisting their separation. This idea has been supported by numerous lattice simulations which have revealed that at large separations the ground state potential increases linearly with the distance $r$ between the quarks (see for example \cite{Bali:1994de}).

In this paper we will model the confining flux tube by generating gauge invariant states $\ket{\Psi}$ which describe possible configurations of the gluonic fields around charges, and testing how well they overlap with the unknown ground state in the mesonic sector, $\ket{\Omega}$. We will build and test various proposals for such a state, using a combination of analytic and lattice techniques. Our focus in this paper will be the cutoff dependence of the models, as imposed by the short distance lattice regulator, and the behaviour of our models in the continuum limit.

Central to the construction of our trial states is the need for gauge invariance: a na\"ive construction of a mesonic state, where we act on the empty vacuum $\ket{0}$ with fermionic creation operators, fails to produce a physical state -- it is not gauge invariant. The origin of the problem lies with the Lagrangian fermions, which themselves are not gauge invariant and do not immediately describe physical observables. In fact, no such local operator can create physical charges in a gauge theory \cite{Lavelle:1995ty}.

The basis of this observation is that charges always appear with an associated (chromo--) electromagnetic field. The result is that the physical electron, for example, is a composite object comprised of a U(1) fermion together with its surrounding electromagnetic fields. This manifestly non-local object is gauge invariant and describes an observable charge \cite{Dirac:1955uv,Lavelle:1995ty, Bagan:1999jf, Bagan:1999jk}. Quantum mechanically, we see that to construct a charged state, or a multi--fermion state with zero overall charge, the fermion creation operators must be accompanied by a configuration of gauge boson creation operators, which generate the necessary chromo--electromagnetic fields. We refer to constructing these fields as `dressing' the charges. Although gauge invariance imposes restrictions on such a state, there is still a huge choice in the form of the dressing. 

It transpires that any gauge fixing condition may be used to define a dressing and therefore a gauge invariant state \cite{Lavelle:1995ty}. Additional conditions on the state can further refine the class of gauge fixing used \cite{Bagan:1999jf}. In particular, the Coulomb gauge is particularly relevant to the description of  static charges and this can be clearly understood perturbatively \cite{Bagan:1999jk}. The extent to which this is also the case non-perturbatively is the focus of this work.

In this paper we concentrate on two model states related to the Coulomb and axial gauge fixing conditions. The `Coulomb state', $\ket{\Phi}$, describes two individually gauge invariant charges each surrounded by non--abelian Coulombic fields. The `axial state', which we denote $\ket{\chi}$ throughout, describes an overall gauge invariant meson formed by linking two fermions by a string of glue. We will formalise these descriptions below. Here we note two properties of these states: first, although our numerical calculations are in SU(2), our analytic constructions apply to SU($N_c$); secondly, if we represent the states as wavefunctionals of (the matter fields and) the gauge field $\A$, then the dressing becomes unity when $\A$ satisfies the gauge condition defining the dressing. For example, the Coulomb dressing may be written as a functional of $\A$ which is always equal to $1$ when $\partial_i A_i=0$. Therefore, if we work in the gauge used to define the dressing, then dressed fermions are just the usual bare fermions.  This observation provides a link between lattice studies of manifestly gauge invariant states and analytical techniques such as the Hamiltonian functional approach based upon Coulomb gauge \cite{Szczepaniak:2001rg, Ligterink:2003hd, Szczepaniak:2003ve, Feuchter:2004mk, Epple:2006hv, Schleifenbaum:2006bq, Alkofer:2005ug}, and investigations of the non--abelian Coulomb potential \cite{Zwanziger:2002sh, Cucchieri:2000gu, Greensite:2003xf, Nakamura:2005ux, Nakagawa:2006fk, Langfeld:2004qs}. 

Once we have constructed a trial state $\ket{\Psi}$ we compare it with the unknown ground state through their (mod--squared) overlap $|\bracket{\Omega}{\Psi}|^2$. This overlap may be calculated on the lattice through the persistence amplitude $\bra{\Psi}e^{-{H}T}\ket{\Psi}$. Since the large Euclidean time limit is a ground state projector, we have
\be\label{persist}
    \bra{\Psi}\e^{-{H}T}\ket{\Psi} \sim e^{-V(r) T}|\bracket{\Omega}{\Psi}|^2\;,\quad T\gg 1\;,
\ee
where the ground state energy in the $q$--$\overline{q}$ sector gives the well--known confining potential, $V(r)$. This holds provided that the overlap between the trial and ground states is non--zero, which has recently been verified for our trial states \cite{Heinzl:2007cp}: the large time persistence amplitudes for both the Coulomb and axial states yield the confining potential $V(r)$, confirming a non--zero overlap with the ground state.

In \cite{Heinzl:2007kx} we studied the overlap for our trial states in U(1) and SU(2) Higgs theories, in both the confining and deconfined phases. In the U(1) theory with heavy charges exact calculations can be performed in the deconfined phase. There we found that the ground state is exactly the Coulomb state $\ket{\Phi}$, describing two individually gauge invariant charges with the familiar electromagnetic Coulombic fields. We found that the axial state $\ket{\chi}$ is an infinitely excited state, giving, in the time slice where it is prepared, a confining potential between the charges. The state is unstable and decays in time to the ground state. Contrastingly, we found that in the confining phase of U(1) the axial state had the larger overlap with the ground state. The physics in this phase therefore appears to be better described by the confining potential generated by a very thin string.

In the Higgs phase of SU(2)--Higgs theory, we found that the Coulomb state had the better overlap with the true ground state. However, contrary to what may have been expected, the Coulomb state also provided a better description of the ground state in the confining phase than the axial state. This implies that the SU(2) flux tube is significantly thicker than the confining string in U(1) theory.

Our previous simulations were performed at a fixed lattice spacing. In this paper we will investigate the dependence of our results on the cutoff and so study the continuum limit. We begin in Section \ref{ax-sect} with the axial state, which describes the flux tube as a string of glue, and comment on some recent results in the literature. In Section \ref{sme-sect} we compare our results with those for a thickened string constructed from smeared links on the lattice. We will see that the overlap between each of these states with the ground state \emph{decreases} as we approach the continuum limit, indicating that they have little to do with ground state physics in the continuum. In Section \ref{cou-sect} we turn to the Coulomb state, which describes a charge--anticharge pair surrounded by non--abelian Coulomb fields. We find  very different behaviour  as the lattice regulator is removed: the overlap of the Coulomb and ground states now \emph{increases} as we approach the continuum. Finally, in Section \ref{concs} we present our conclusions.

\section{The axial state}\label{ax-sect}
\subsection{Construction and properties}
The axial state is
\be\label{ax-state}
	\ket{\chi}=\overline{q}(\x_2)\, \text{P}\exp\bigg[\int\limits_{\x_2}^{\x_1}\!\ud z_j{A}_j(\z)\bigg]\, q(\x_1)\ket{0}\;,
\ee
with $q(\x)$ the heavy fermion source.  This state describes a gauge invariant, charge neutral meson formed by linking the matter fields by a gluonic string. We take the string to lie on the straight line connecting $\x_1$ and $\x_2$, and so $r\equiv |\x_2-\x_1|$ is the separation of our fermions. For the axial state, calculating the persistence amplitude over time $T$ corresponds to calculating an ordinary rectangular (and unsmeared) Wilson loop of spatial extent $r$ and temporal extent $T$ \cite{Heinzl:2007cp}. Before presenting the numerical results it will be useful  to outline the behaviour of the analogous state in QED with heavy charges. There the ground state is known exactly, and its overlap with the axial state is zero \cite{Heinzl:2007kx}. The origin of this vanishing overlap can be traced back to short distance effects, as the flux is trapped on an infinitely thin string, even though the axial wavefunctional contains no UV (nor IR) divergence. Including a momentum cutoff $\Lambda$, the overlap is found to be, up to terms finite as $\Lambda\to\infty$ in the exponent,
\be\label{u1over}
	|\bracket{\Omega}{\chi}|^2 = (r\Lambda)^{4\alpha/\pi}\exp\big[-\alpha\, r\Lambda+\ldots\big]\;.
\ee
Here $\alpha=e^2/(4\pi)$. The overlap is a function of $r \Lambda$, dominated by exponential decay for large values of the cutoff which probe the infinitesimal extent of the string in the two directions transverse to $\x_2-\x_1$. As the cutoff is removed this causes the overlap with the ground state to vanish.
 
We now describe our calculation of the unsmeared SU(2) Wilson loop. We used a standard heat bath algorithm combined with microcanonical reflections to generate an ensemble of 1000 configurations which were used to measure the persistence amplitudes.  Defining a supersweep to be a combination of 3 heat bath update sweeps followed by 7 microcanonical reflections to enhance the ergodicity of the algorithm, we allowed 250 supersweeps to reach thermal equilibrium. We found no dependence of our results on the initial configuration chosen for the lattice. All measurements were taken on a 20$^4$ lattice, of varying spacings $a$, using a series of configurations separated by 10 supersweeps. In analysing our data we ruled out any measured value of the persistence amplitudes with relative error larger than 0.5. The persistence amplitude data was fitted to the formula
\be\label{persist2}
    \bra{\Psi}\e^{-{H}T}\ket{\Psi} = e^{-V(r) T}|\bracket{\Omega}{\Psi}|^2 \;,
\ee
which assumes that the elapsed time $T$ is large enough for contributions from excited states to decouple, following~(\ref{persist}). The axial state data we present was derived from measurements taken at an elapsed time of $T\geq5a$, for which no significant deviations from (\ref{persist2}) were observed. The overlaps were extracted from the logarithm of the data by a weighted least-squares fit to a straight line (measured for a fixed value of spatial separation $r$ of the static charges). The quality of such a fit can be controlled by inspection of the values of two parameters: $Q$, the goodness-of-fit, and the reduced $\chi^2$ value, $\chi^2/ \nu$ where $\nu=N-2$ denotes the number of degrees of freedom for a linear fit of $N$ data points (this is not to be confused with the state $\ket{\chi})$.  Our acceptance criteria were  $\chi^2 / \nu < 3$ and $Q> 10^{-3}$.

\subsection{Numerical results}
\begin{figure}[t!]
\includegraphics[width=8cm]{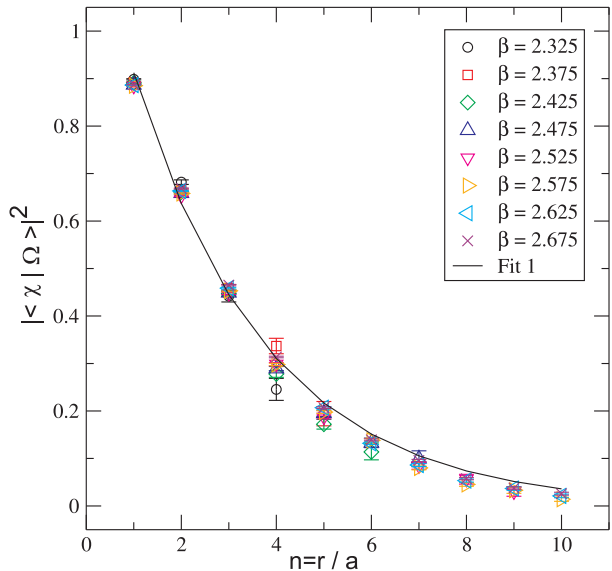}
\caption{Overlap between the axial state and the ground state, at various lattice spacings, plotted against the number of lattice points $n=r/a$ between charges (for simulation details, see the text). The overlap is independent of $\beta$.}
\label{fig:axial1}
\end{figure}
\begin{figure}
\includegraphics[width=8cm]{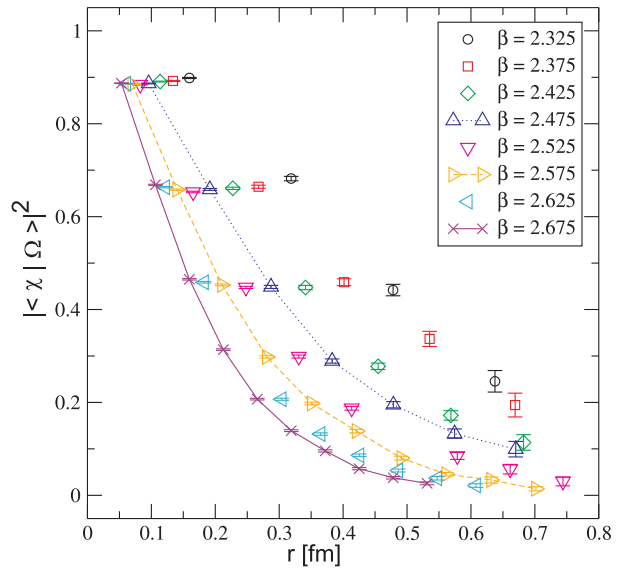}
      \caption{Overlap between the axial state and the ground state, plotted against the physical separation of the charges. Some lines have been added to guide the eye. It is seen that, for all $r$, the overlap is smaller for finer lattice spacings (larger $\beta$).}
\label{fig:axial2}
\end{figure}
\begin{figure*}
\includegraphics[width=11cm]{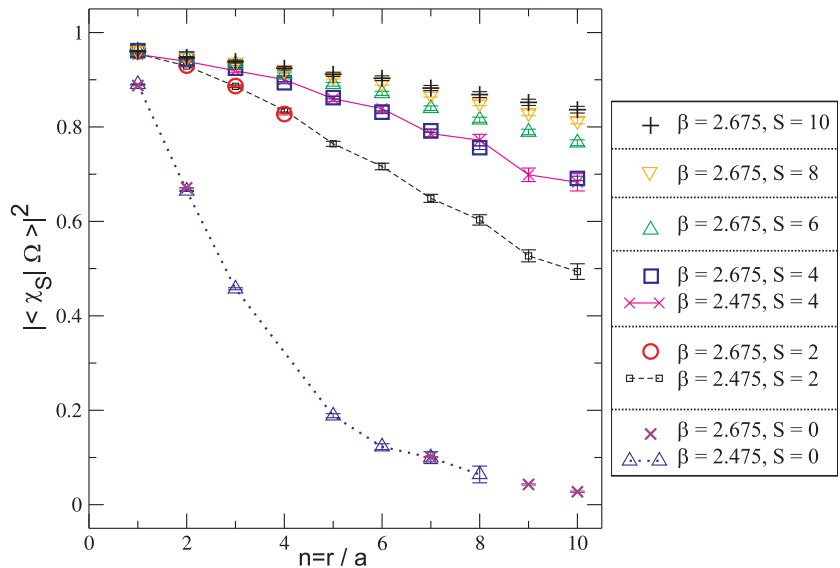}
\caption{Smeared state overlap with the ground state at various lattice spacings. For a fixed smearing level $S$ the results lie on the same curve, they are $\beta$ independent. We have added data for the axial state ($S=0$) for comparison, and some lines to guide the eye. Lattice: $500$ configurations, $T=4a$, relative error of Wilson line correlator $<0.5$, $\chi^2/\nu<3$ and $Q>10^{-2}$. }
\label{fig:smear1}
\end{figure*}
In Figure \ref{fig:axial1} we plot the overlap of the axial and ground states against the number of lattice sites $n$ between the charges. Only data points for the overlaps with relative errors smaller than 0.35 are displayed. We note first that all data points lie on top of one another, signalling that the overlap is independent of the lattice spacing (and therefore $\beta$) used for the simulations. Given that the U(1) overlap, (\ref{u1over}), is a function of $r\Lambda$, we may expect that the SU(2) overlap is a function of $r/a$, as $1/a$ is the UV cutoff on the lattice. We therefore fit the data to an exponentially decaying function of the form
\be\label{fitforms}
|\bracket{\Omega}{\chi}|^2= C \exp\bigg(-\lambda\,\frac{r}{a} \bigg)\;.
\ee
The fit is shown as the solid line in Figure \ref{fig:axial1}. We observe a very close fit to the data, for the parameter values $C=1.307$, $\lambda=0.359$.

For any given charge separation $r$, the continuum limit $a\to0$ corresponds to $n\to\infty$. Our simulations clearly show us that as $n$ increases the overlap of the axial state with the ground state drops exponentially. Both this exponential decay and the dependence on $r/a$ parallel the U(1) results.  As we approach the continuum limit, $a\to0$ (which corresponds to $\beta \to \infty$), we probe more of the ultraviolet artifacts of the infinitesimally thin string which lowers the overlap with the ground state. 

The overlap may also be expressed as a function of the physical separation of the charges. The conversion was performed by interpolating between known values for the string tension $\sigma a^2(\beta)$ at various values of $\beta$. The string tensions and the perturbative 1--loop interpolation formula were both taken from \cite{Langfeld:2007zw}. Figure \ref{fig:axial2} displays the overlap as a function of $r$ at various $\beta$. As $\beta$ increases the overlap decreases for all separations of the charges. If we fix $r$ at a given separation and take the continuum limit, $\beta\to\infty$, the overlap again tends to zero. If the trends observed in our results continue to hold at smaller lattice spacings, then in the continuum limit the overlap with the ground state will vanish.

Recent papers \cite{Boyko:2007ae, Boyko:2007jx} have made two claims regarding the geometry of the SU(2) flux tube: firstly, that at a fixed lattice spacing the width of the flux tube grows with increasing charge separation but, secondly, that at finer lattice spacings thickening of the tube is a subleading effect, as the width of the tube is proportional to the lattice spacing.  The first claim agrees with the interpretation of our above results -- we observe that the very thin string is a poorer description of the ground state at larger separations. The second claim would imply that in the continuum limit all  thickening of the tube is suppressed, and that the flux tube of SU(2) would in fact be an infinitely thin string of flux. However, we have seen that for a given $r$ the overlap of this axial state with the ground state drops exponentially as we move toward the continuum limit.

The linear $a$ dependence of the flux tube width described in \cite{Boyko:2007ae, Boyko:2007jx} was observed to set in at around $a\simeq 0.06$\,fm, or $\beta\simeq 2.600$. Our largest $\beta$ values ($\beta=2.675$) probe this region, yet we see no change to the functional form of exponential decay. We predict that in the continuum limit the overlap between the axial and ground states vanishes, and that the SU(2) ground state is \emph{not} well described by a thin string stretched between the charges.

\section{Smeared states}\label{sme-sect}
\subsection{Smearing the string}
If a thin string is not a good description of the $q$--$\overline{q}$ ground state of the glue, then we must look for a state which describes a more dispersed distribution of glue. We have seen in U(1) theory that it is the infinitesimal thickness of the string which leads to a vanishing overlap with the ground state, and a similar interpretation is supported by our numerical results in SU(2). As thickening the string will soften the UV behaviour we would therefore hope to see an improvement in the overlap with the ground state.

A state with a thicker string may be prepared on the lattice using ``smearing" \cite{Teper:1987wt, Bali:1992ab,  Albanese:1987ds}. Smearing replaces links by a sum of their adjacent staples projected onto an SU(2) group element. The smeared axial state is prepared by smearing the links between the charges, the effect being to broaden the string of glue. It is well known that the use of such smeared operators greatly improves the accuracy of calculations of, for example, the glueball spectrum \cite{Morningstar:1999rf} and the inter--quark potential \cite{Juge:2002br}. This is because smearing an operator reduces its sensitivity to higher excitations of the theory, improving the projection onto the ground state in the calculation of large time Wilson loops.  This is the origin of the alternative name `overlap enhancement'. We expect, then, that removing the UV modes from our state by smearing the string should give us a state which is closer to the true ground state.

\subsection{Numerical results}

Repeated smearing gives an increasingly smoother and more dispersed configuration. We smeared our axial state with a number $S$ of smearing steps from $S=1$ to $S=10$. Wilson loops with $T\geq 4a$ were sufficient to project onto the ground state and 500 lattice configurations were used in the simulations.

\begin{figure}
\includegraphics[width=8cm]{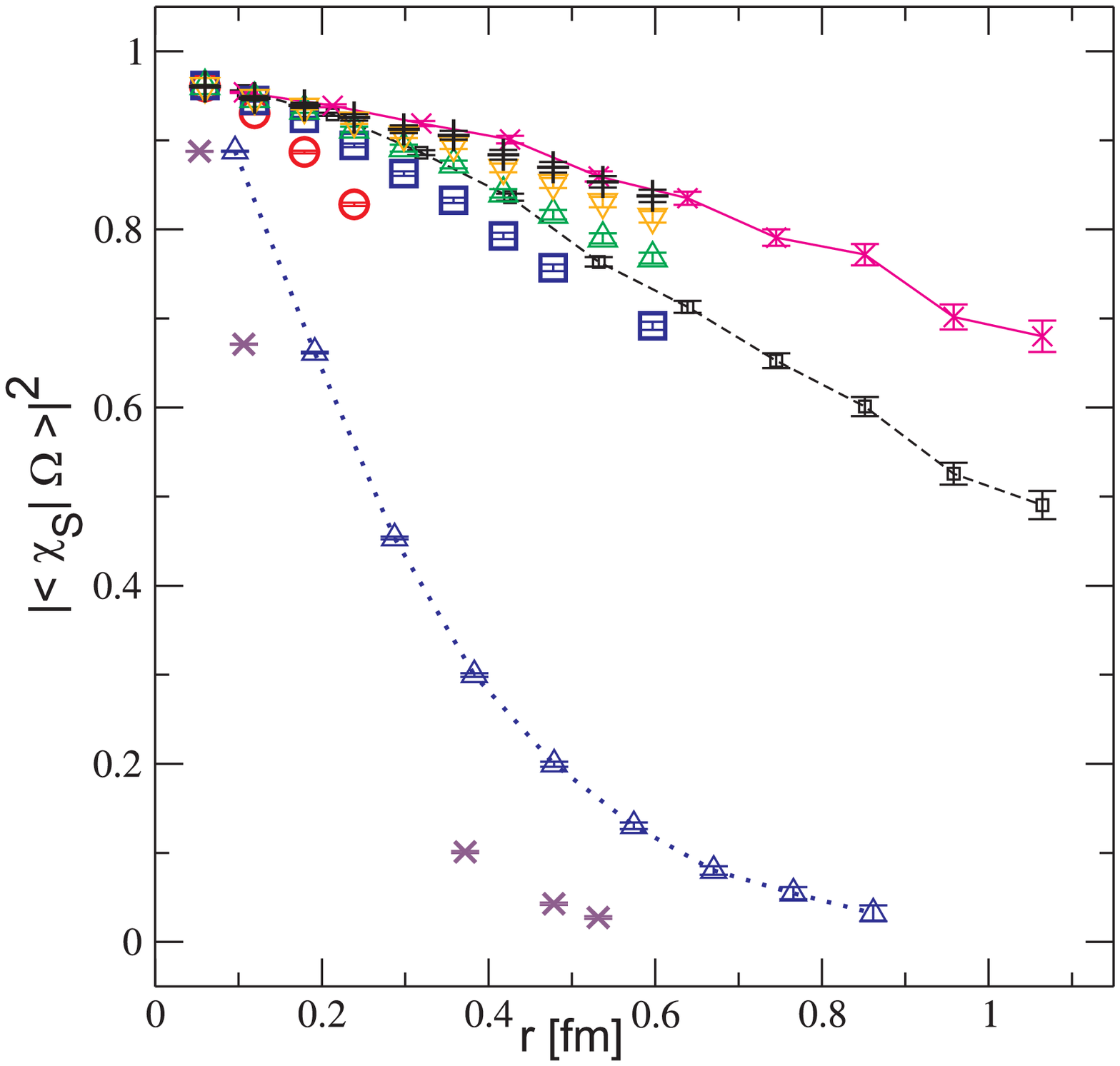}
\caption{Smeared state overlap with the ground state, plotted against physical separation. Axial state data ($S=0$) is displayed for comparison and some lines have been added to guide the eye. Legend as in Figure \ref{fig:smear1}.}
\label{fig:smear2}
\end{figure}

Our results are plotted in Figure \ref{fig:smear1}, as a function of $r/a$, and in Figure \ref{fig:smear2} as a function of the separation $r$. These plots also show some axial state data (no smearing steps, i.e. $S=0$) for comparison. For a given separation, smearing clearly improves the overlap with the ground state. Figure \ref{fig:smear1} shows us that the overlap is $\beta$ independent at fixed $S$ (as it was for the axial state, $S=0$). For a fixed $r$, we approach the continuum limit by following one of the curves in Figure \ref{fig:smear1} to the right. Each curve falls as we go to the continuum so the overlap drops to zero. Although not as severe as the exponential drop of the axial ($S=0$) state, the overlap at a given $r$ still decreases as we go to finer lattices, suggesting that the smeared overlap also vanishes in the continuum limit.

The overlap enhancement provided by a given $S$ is therefore fixed and does not increase as we go to finer lattices.  The practice of smearing to, say, 10 steps, is an artifact of our current lattice sizes. As computations on finer lattices become more common, a given level of smearing will not produce the same improvement in the overlap as for current lattices. Instead, the smearing level will have to be increased to see a comparable improvement.

So far, we have considered states which describe a thin string, or a thicker string produced from smearing. We have seen that the thin string gives an increasingly worse description of the ground state as we go to the continuum limit. We have also seen that the amount of overlap enhancement provided by smearing depends not on $\beta$ but just on the number of smearing steps. The overlap of the smeared state also tends to zero in the continuum limit.

Higher numbers of smearing steps describe successively more widely spread gluonic fields, and increasing the number of smearing steps increases the overlap with the ground state (at a given $\beta$). We will therefore now turn to a model which we expect to describe more widely distributed glue. We will find a markedly different behaviour of the overlap in the continuum limit.

\section{The Coulomb state}\label{cou-sect}

\subsection{Construction and properties}
The axial state (\ref{ax-state}) describes a heavy meson without gauge invariant constituents. Dressings can, though, also give us individually gauge invariant colour charges, describing the expected short--distance physics of QCD. We will now examine the dressing which generalises Dirac's static electron \cite{Dirac:1955uv} to a static, heavy quark in non-abelian gauge theories \cite{Lavelle:1995ty}. A gauge invariant charge can be created from the vacuum by the operation
\be
	h^{-1}[\A,\x]q(\x)\ket{0}\;,
\ee
where $h^{-1}$ is a functional of $\A$ obeying \cite{Lavelle:1995ty, Ilderton:2007qy}
\be\label{use}
	\partial_j\bigg( h^{-1} A_j h+\frac{1}{g}h^{-1}\partial_j h\bigg)=0\;.
\ee
This equation is to be understood in a Schr\"odinger representation where we have diagonalised the operator $\A$. It defines $h[\A,\x]$, the field dependent transformation which takes $\A$ into Coulomb gauge. The dressing itself transforms under gauge transformations as $h^{-1}\to h^{-1}U$ if $q\to U^{-1}q$, giving a gauge invariant fermion. Equation (\ref{use}) may be used to perturbatively construct this Coulombic dressing by expanding in powers of the coupling, giving a gauge invariant charge with a well defined colour \cite{Lavelle:1995ty, Ilderton:2007qy}. A dramatic simplification occurs if we choose to work in Coulomb gauge, however, as the dressing becomes a trivial factor of unity in colour space \cite{Lavelle:1995ty, Ilderton:2007qy}, i.e.
\be
	h^{-1}[\A,\x]\bigg|_{\partial_i A_i=0} = 1\;.
\ee
This is easy to see from the definition of $h[\A,\x]$ as the rotation into Coulomb gauge -- if we are already in that gauge, no rotation is needed. Our non-abelian, colour singlet Coulomb state $\ket{\Phi}$ contains two such gauge invariant charges separated by a distance $r$. In Coulomb gauge we therefore have
\be
	\ket{\Phi} = \overline{q}(\x_2) q(\x_1)\,\ket{0}\;,
\ee
traced over colour indices. In perturbation theory, the gluonic fields around these charges are distributed over all space, and the potential $\bra{\Phi}H\ket{\Phi}$ between the charges is, at lowest order,
\be
	\bra{\Phi}H\ket{\Phi} = -\frac{g^2 C_F}{4\pi}\frac{1}{r}+ \text{self energies}+\mathcal{O}(g^4)\;.
\ee
This is the lowest order contribution to the familiar interquark potential. At higher orders, screening and anti--screening structures emerge \cite{Bagan:2005qg}. It is useful here to outline the properties of the corresponding state in U(1). There, if we have only heavy charges, the Coulomb state is the ground state in the charge--anticharge sector. It describes two individually gauge invariant charges surrounded by Coulomb fields. When we allow light fermions, however, we know that the effect of virtual pairs is to screen the charge, which lowers the energy. This results in an additional, but gauge invariant, contribution to the U(1) Coulomb state which incorporates the screening effects \cite{Bagan:2001wj} and has the form of Polyakov lines.  Returning now to the non--abelian theory, it is well known \cite{Drell:1981gu} that glue both screens and anti--screens the heavy charges. The dressing $h^{-1}[\A,\x]$ contributes the anti--screening effects \cite{Lavelle:1998dv}. The additional, gauge invariant contributions which give screening effects come from short Polyakov lines generated as the state evolves in time.

Our Coulomb dressing  describes individually gauge invariant charges in perturbation theory, where we can describe the physics of our dressings analytically. Non--perturbatively we know that any description of a physical colour charge must break down, as no such objects are observed (as asymptotic states) in nature.  The Gribov ambiguity has been shown to generate just this breakdown \cite{Lavelle:1995ty,Ilderton:2007qy}. Confinement will appear at the scale where Gribov copies reintroduce an unphysical gauge dependence to the single charge dressing. Beyond this scale only dressings for hadronic (colourless) states exist. This could force our two individual dressings, $h[\A,\x_2]$ for $\overline{q}(\x_2)$ and $h^{-1}[\A,\x_1]$ for $q(\x_1)$, to combine into a single mesonic dressing, i.e., the Coulomb state could still, non--perturbatively, confine the charges \cite{Heinzl:2007cp, Ilderton:2007qy}. In fact, it is known from lattice results that the non--abelian Coulomb gauge potential, which is an upper bound to the full potential, is confining with a linear rise \cite{Zwanziger:2002sh, Greensite:2004ke}, in contrast to the abelian theory.

We now turn to the numerical preparation of our state and the calculation of its overlap with the ground state. We will see that, non--perturbatively, the Coulomb state is a good approximation to the true ground state, and that their overlap in fact \emph{improves} as the lattice regulator is removed. 
\begin{figure}
 \includegraphics[width=8cm]{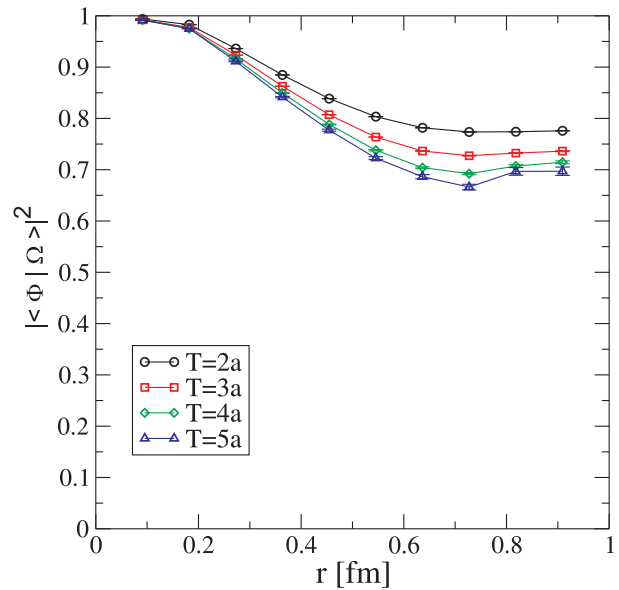}
\caption{Improved action simulation of the Coulomb overlap at fixed $\beta=1.55$, $20^4$ lattice. As $T$ increases contributions from excited states decouple, and we see a decrease in the overlap, which nevertheless remains above 0.7 for the $T$ values plotted -- however, the difference between results at different $T$ is decreasing.}
\label{fig:Coul1}
\end{figure}

\begin{figure}
 \includegraphics[width=8.5cm]{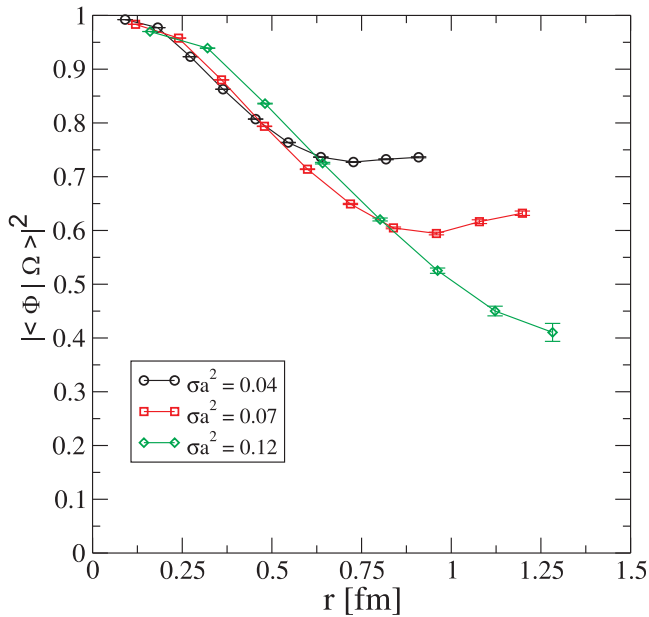}
\caption{Improved action simulation of the overlap between the Coulomb and ground states. At a fixed time, here $T=3a$, we observe an increase in the overlap in the continuum limit ($\beta\to\infty$ or $a\to0$).}
\label{fig:Coul2}
\end{figure}
\begin{figure}
 \includegraphics[width=8.5cm]{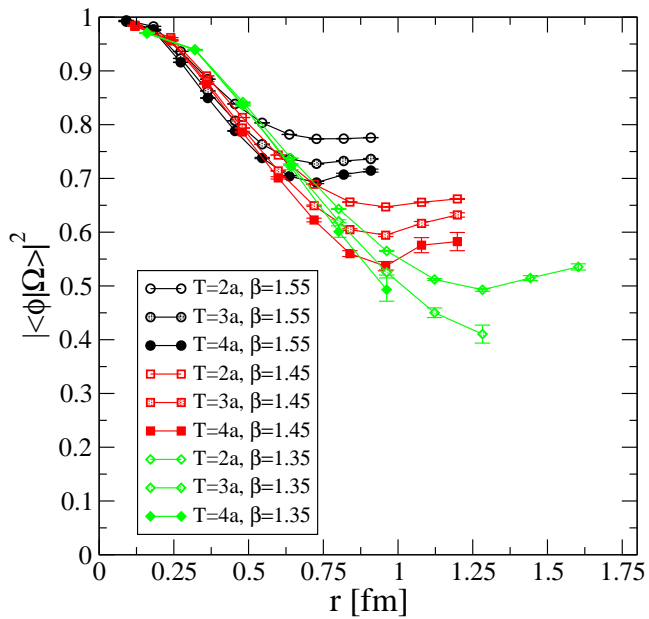}
\caption{A summary of the improved action results for the Coulomb overlap, plotted at a range of $\beta$ and $T$ values.}  
\label{fig:Coul3}
\end{figure}

\subsection{The Coulomb state on the lattice}

Our simulations were performed using an improved lattice action \cite{Langfeld:2007zw} which gives good scaling and rotational symmetry properties.  Calculating the persistence amplitude for the Coulomb state requires bringing the lattice into Coulomb gauge and evaluating correlators of Polyakov lines in the gauge fixed configuration \cite{Heinzl:2007cp}. We used an iteration--overrelaxation algorithm to fix the Coulomb gauge by maximising the gauge fixing functional
\be \label{Coulomb Gauge Functional}
 \mathcal{F}[U^\Omega] = \frac{1}{2}\ \Re \, \bigg\{ \tr \bigg[ \sum_{x}
\sum^{3}_{i=1} U^\Omega_{i}(x) \bigg] \bigg\} \;.
\ee
During gauge fixing we monitored the behaviour of
\be
\delta^2 = \frac{1}{4 N} \sum_{x=1}^{N}\sum_{b=1}^{3} \big(\tr[\tau^{b} B(x)]\big)^{2}\;.
\ee
Here $B(x) = \sum_{i=1}^{3}[ U_{i}(x) + U^{\dagger}_{i}(x- \hat{e}_{i})]$ and $\tau^{b}$ are the three generators of SU(2).
Expanding in the lattice spacing we find
\be
\tr[\tau^{b} B(x)] = a^{2} \partial_{i} A^{b}_{i}(x) + \mathcal{O}(a^{3})\;,
\ee
which shows that $\delta^2$ is a measure of violation of the Coulomb gauge condition. As a stopping criterion we demanded $\delta^{2} < 10^{-10}$.  Our simulations cover quark separations from roughly 1.8 fm down to 0.05 fm.

In a first run, we studied the $T$ behaviour of the data using a $20^4$ lattice and $\beta=1.55$, with the aim of identifying $T_\text{min}$ such that excited states effectively decouple for $T\geq T_\text{min}$. The results of this run are shown in Figure~\ref{fig:Coul1}, which plots the overlap as a function of $r$ for several values of $T_\mathrm{min}$.

Although the static  potential is stable for $T_\mathrm{min}\ge 3$, the overlap retains a $T_\mathrm{min}$ dependence.  The high value of the overlap, over $70\%$, and the sensitivity to $T_\text{min}$ indicate that the Coulomb state has a significant overlap not only with the ground state but also with low lying excited states, which have not decoupled. We must therefore go to larger $T$ values to see the overlap with only the ground state. It seems, though, that the higher modes will decouple not far beyond $T=5a$, as our results show that the \emph{difference} between overlaps decreases as $T$ is increased, and the curves at $T=4a$ and $T=5a$ are already very close together. The results at our larger $T$ values should therefore serve as a good approximation to the true overlap. Having addressed the residual $T$--dependence, we now study the overlap for varying $\beta$. We note first, however, that the dependence on $T_\mathrm{min}$ suggests a small or possibly zero mass gap in the quark--antiquark channel. 

We plot in Figure \ref{fig:Coul2} the overlap at various lattice spacings and fixed $T$. Figure \ref{fig:Coul3} summarises the data, showing both the $T$ and $\beta$ dependencies. The difference between the behaviour of the Coulomb and axial states in the continuum limit is striking -- for a given $r$ the Coulomb overlap increases with decreasing lattice spacing, becoming a better description of the true ground state as we approach the continuum. Does this pattern hold as we go to finer lattices? To answer this question and study our Coulomb state closer to the continuum limit we now turn to the Wilson action, which will allow us to study the overlap at larger $\beta$ values. This will also show that our results are independent of the discretisation method employed. We present the results of these simulations below.

\subsection{Wilson action results}
\begin{figure}
 \includegraphics[width=8.6cm]{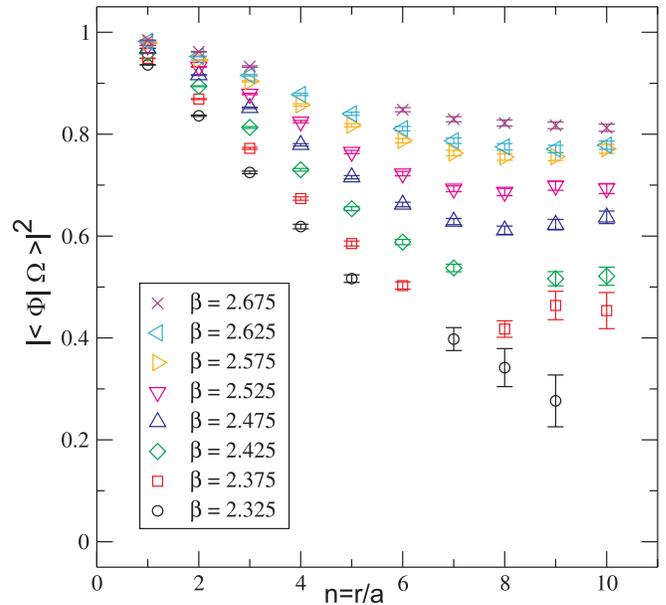}
\caption{Wilson action overlap of the Coulomb state and ground state, plotted against the number of lattice sites $n=r/a$ between the charges. For lattice data, see the text. Measurements were taken from $T\geq 3a$. Notably, the overlap increases as we go to the continuum, $\beta\to\infty$.}
\label{fig:Coul4}
\end{figure}
\begin{figure}
\includegraphics[width=8.6cm]{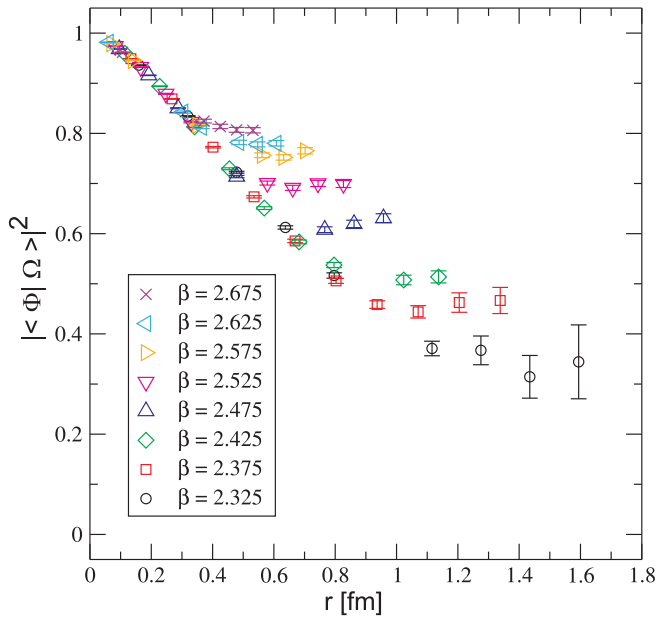}
\caption{Wilson action overlap of the Coulomb state and ground state, plotted against the physical separation of the charges. The overlap increases as we increase $\beta$.}
\label{fig:Coul5}
\end{figure}

Our Wilson action results for the Coulomb overlap are shown in Figure \ref{fig:Coul4} as a function of $r/a$. We observe a markedly different behaviour to both the axial and smeared states. Firstly, we note that the results do not all lie on the same curve. The form of the data no longer fits a $\beta$ independent function of $r/a$, as found for the axial overlap, Figure \ref{fig:axial1}. Instead, as $\beta$ increases the Coulomb overlap \emph{increases} for all values of $r/a$. The origin of the axial overlap's dependence only on $r/a$ was the UV physics of the infinitesimally thin string of glue between the charges. The configuration of glue described by the Coulomb dressing has no such structure. This eliminates the UV artifacts of the axial state and so the Coulomb overlap depends on both $r$ and $a$ separately rather than on only their ratio.  

We now turn to the overlap as a function of the physical separation of the charges, plotted in Figure \ref{fig:Coul5}, analogously to Figure \ref{fig:Coul2}. Notice that the order of the curves corresponding to simulations at different lattice spacings is reversed as compared to the axial plots.   At short separations the overlap with the ground state is almost perfect and drops more slowly with increasing $r$ than the exponential drop of the string state (see Figure \ref{fig:axial2}). It is also almost independent of the lattice spacing in this regime.  As we draw the two charges apart, our physical expectation is that a flux tube forms, confining the charges into a meson. In this situation the Coulombic description may not be expected to be appropriate, and indeed we see that the overlap drops with increasing distance. However, this trend does not continue indefinitely. For a given lattice spacing we see, at some physical separation, a levelling off in the overlap. This phenomenon occurs at different distances for different lattice spacings -- as the lattice spacing is decreased, and we move to the continuum limit, this levelling of the overlap begins at smaller and smaller physical separation, giving a good and almost $r$ independent overlap with $\ket{\Omega}$. The Wilson action results therefore confirm our previous findings with the improved action, Figure~\ref{fig:Coul3}.

The most significant difference between the axial and Coulombic states is that the overlap of the former (and its smeared counterparts) decreases as we go to the continuum limit, for all separations of the charges. The Coulombic state, however, has a good overlap with the ground state which \emph{increases} in the continuum limit. For any charge separation $r$, we move up the graph in Figure \ref{fig:Coul5} away from $0$ and towards $1$.

The small difference from $1$ in the overlap of the Coulomb and ground states is reminiscent of perturbative results in (Coulomb gauge) time--independent perturbation theory \cite{Drell:1981gu, Lee:1981mf}. The overlap between the $g=0$ ground state (essentially $N_c^2-1$ copies of the abelian theory) and the perturbed $q$--$\overline{q}$ sector ground state goes like $1-\mathcal{O}(g^2)$ for $g$ small. Here we have seen that the overlap between our Coulomb state and the true ground state differs, non--perturbatively, by a small quantity which decreases as we go to the continuum.  We therefore expect that the physics of the Coulomb state should give a good description of the true ground state physics, implying in particular that the ground state of the glue around heavy charges takes a form much thicker than that of a thin string. Due to non--perturbative effects, such as the Gribov ambiguity, we cannot rule out the possibility that, non--perturbatively, the Coulomb state becomes the true ground state in the continuum limit -- as discussed, the non--abelian Coulomb potential is linearly rising for large separations and a good agreement between the Coulomb string tension and the full string tension was reported in~\cite{Langfeld:2004qs}. 

\section{Conclusions}\label{concs}
We have constructed and analysed models of the $q$--$\overline{q}$ sector ground state in Yang--Mills theory. Our states are gauge invariant and describe  different distributions of glue around the fermions.

Our first ansatz, the axial state, modelled the flux tube by a string of glue stretched between the fermions. We calculated the overlap between this state and the ground state in SU(2). We saw that for any separation of the fermions the overlap decayed exponentially with the ultraviolet cutoff provided by the inverse lattice spacing. These results parallel those of U(1), where the infinitesimal transverse extension of the string leads to a vanishing overlap between the axial state and the ground state in the continuum. We conclude that the SU(2) flux tube is not well described by a very thin string.

This should be contrasted to the confining phase in compact U(1) where, at a fixed lattice spacing, we have seen \cite{Heinzl:2007kx} that the axial state provides a good description of the ground state.  Investigating whether this statement holds in the continuum limit could shed light on the differences between abelian and non--abelian confinement (see also \cite{Shifman:2008zr}). It is also interesting to speculate on the role of the SU($N_c$) axial state for large $N_c$. Here Yang--Mills theory is expected \cite{'t Hooft:1973jz} to have a dual description as a theory of strings, and it may be that in such a limit the axial state again becomes a better description of the ground state. 

Smeared states are formed on the lattice by replacing links with sums over staples. The effect of this procedure is to broaden the string, removing the UV modes. We found that the overlap enhancement offered by smearing is independent of $\beta$ for a given number of smearing steps. Our results show that, for a given separation of the charges, the overlap between the smeared state and the ground state still drops toward zero in the continuum limit, though it does so more slowly than for the unsmeared axial state. This implies that for larger lattices more and more smearing will be required to improve simulations.

Our final model was the Coulomb state, which describes, in perturbation theory, two individually gauge invariant colour charges surrounded by non--abelian Coulombic fields. This ansatz has through its relationship to Coulomb gauge fixing \cite{Lavelle:1995ty,Ilderton:2007qy} and its Gribov copies, a rich non--perturbative structure. We observed that it has a much better overlap with the ground state than the axial descriptions did, which implies it captures more of the true ground state physics. Since the overlap actually increases as the lattice spacing is reduced, we cannot exclude the possibility that the Coulomb state is indeed the true ground state of the $q$--$\overline{q}$ sector in the continuum limit.

There exist many further questions to explore -- for example, it would be interesting to study the role of Gribov copies in our numerical calculations and their contributions as a function of the lattice spacing.  The result that the Coulombic description of the ground state yielded the best overlap at small lattice spacings deserves future study, using large scale simulations to further approach the continuum. Our understanding of the geometry of the flux tube would be increased by constructing trial states with a transverse profile for the tube and maximising their overlap with the ground state. It would be also be very interesting to explore our surprising result that the data supports a small, or zero, mass gap in the meson sector.

\vspace{-20pt}\acknowledgments
The authors thank Andreas Wipf for very useful discussions. The numerical calculations in this paper were carried out on the HPC and PlymGrid facilities at the University of Plymouth.

\end{document}